Universidade Federal de São Carlos

Centro de Ciências Exatas e de Tecnologia

Departamento de Computação

# A Tutorial on Principal Component Analysis with the Accord.NET Framework

César R. Souza

cesar.souza@dc.ufscar.br

16th May, 2012

**Abstract**

This documents aims to clarify frequent questions on using the Accord.NET Framework to perform statistical analyses. Here, we reproduce all steps of the famous Lindsay's Tutorial on Principal Component Analysis, in an attempt to give the reader a complete hands-on overview on the framework's basics while also discussing some of the results and sources of divergence between the results generated by Accord.NET and by other software packages[1].

---

[1] This paper is under constant development; any suggestions can be sent to the author at the given electronic address.

# Contents

Please cite this document as:

# Code listings

A package containing all source codes referenced in this paper is available from
http://www.dc.ufscar.br/~cesar.souza/publications/pca-tutorial.zip.

# 1 Introduction

Principal Component Analysis (PCA) is a technique for exploratory data analysis with many success applications in several research and application fields. It is often used in image processing, data analysis, data pre-processing and visualization. One of the most popular resources for about PCA is the excellent tutorial due to Lindsay I Smith [1]. On her tutorial, Lindsay gives an example application of PCA, detailing all steps involved. While we will be leaving out the more detailed discussions about the exposed concepts to Lindsay and not address them in detail in this text, we will be presenting some practical examples to reproduce most of the calculations given in her tutorial.

The Accord.NET Framework [2] provides methods and algorithms on machine learning, mathematics, statistics, computer vision, computer audition, and several other scientific computing techniques to the .NET environment. The project extends the popular AForge.NET Framework providing a more complete scientific computing suite. The framework has been used in a number of scientific publications [3,4,5,6,7] and academic works [8,9,10,11,12] with overall success.

First and foremost, to follow this guide it is necessary to have the Accord.NET Framework installed. Instructions for download, installation, and sample project creation are available at the project page[1]. Packages are also available through NuGet [13]. In order to use the Extension methods presented in this tutorial, it is necessary to import the *Accord.Math* and *Accord.Statistics* namespaces in your source code.

---

[1] Hosted at: http://accord.googlecode.com

# 2  Background Mathematics

Reflecting much of the original work by Lindsay, we will start by exposing some basic mathematical concepts which will be fundamental in the understanding of Principal Component Analysis. This section will also present code examples on how to perform basic matrix manipulation and simple statistic calculations within the framework.

In order to use the Extension methods presented in this tutorial, it is necessary to import the *Accord.Math* and *Accord.Statistics* namespaces in your source code. To do so, on top of your class source code, add using directives for the namespaces *Accord.Math* and *Accord.Statistics*. Other code examples may also require the *Accord.Math.Decompositions* namespace.

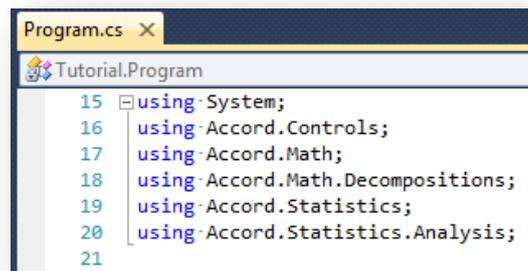

**Figure 1.** Some of the required namespaces.

The Accord.NET uses C# 3.0 extension methods to offer many of the standard operations expected from a Matrix library. Those operations include linear system solving, matrix algebra and numerical decompositions – all of them directly applicable to standard .NET arrays and matrices.

## 2.1 Statistics

This section shall address some basic concepts of statistics.

### *2.1.1 Mean*

The mean is a measure of location. The sample mean gives an estimate of the expected value for a given population. It is not to be confused with other measures such as the median and the mode. For a data set, the mean is simply the arithmetic average of its numbers. For demonstration, here we will use the same set of numbers used by Lindsay, pg. 3.

$$X = [\ 1\ 2\ 4\ 6\ 12\ 15\ 25\ 45\ 68\ 67\ 65\ 98\ ]$$

Now, we would like to compute $\bar{X}$, the mean value of $X$. First, we should express X as a .NET array. To do so, we declare

```
double[] X = { 1, 2, 4, 6, 12, 15, 25, 45, 68, 67, 65, 98  };
```

Suppose we would like to compute the average of those numbers by first computing its sum and then dividing by the number of samples. One approach is to use

```
double x̄ = X.Sum() / n;
```

in which n is set to be `X.Length`. A more direct approach is to add an using directive to the *Accord.Statistics* namespace and then use

```
double x̄ = X.Mean();
```

In both cases, the result of those operations stored on x̄ should be 34.

*2.1.2 Standard Deviation*

Often, the mean alone does not give us much information about the data we have – it is not a sufficient description. The mean only gives us a sense of location, but doesn't tell us how far individual data points spread from this location. The Standard Deviation, on the other hand, measures spread, and as such, gives us an idea about the dispersion of the data. Lindsay gives us an example on how two data sets can have the same mean, but different standard deviations. Consider the two sets with the same mean (pg. 3)

```
double[] x1 = { 0, 8, 12, 20 };
double[] x2 = { 8, 9, 11, 12 };
```

First we should compute their mean, making sure the two sets have indeed the same mean:

```
double mean1 = x1.Mean();
double mean2 = x2.Mean();
```

Now let's compute their standard deviation:

```
double stdDev1 = x1.StandardDeviation();
double stdDev2 = x2.StandardDeviation();
```

Since we already have pre-computed the mean, we can avoid repeating some calculations by passing the mean into the Standard Deviation computation:

```
double stdDev1 = x1.StandardDeviation(mean1);
double stdDev2 = x2.StandardDeviation(mean2);
```

And this is all that is needed to compute the mean and standard deviations of two data samples. The code listing in the next page summarizes how to compute the mean and standard deviation of those two data sets.

| Program | Output |
|---|---|
| ```csharp
// Create some sets of numbers
double[] x1 = { 0, 8, 12, 20 };
double[] x2 = { 8, 9, 11, 12 };

// Compute the means
double mean1 = x1.Mean();
double mean2 = x2.Mean();

// Compute the standard deviations
double stdDev1 = x1.StandardDeviation(mean1);
double stdDev2 = x2.StandardDeviation(mean2);

// Show results on screen
Console.WriteLine("Data:");
Console.WriteLine(" x1: " + x1.ToString("G"));
Console.WriteLine(" x2: " + x2.ToString("G"));
Console.WriteLine();

Console.WriteLine("Means:");
Console.WriteLine(" x1: " + mean1);
Console.WriteLine(" x2: " + mean2);
Console.WriteLine();

Console.WriteLine("Standard Deviations:");
Console.WriteLine(" x1: " + stdDev1);
Console.WriteLine(" x2: " + stdDev2);
``` | ```
Data:
 x1: 0 8 12 20
 x2: 8 9 11 12

Means:
 x1: 10
 x2: 10

Standard Deviations:
 x1: 8.32666399786453
 x2: 1.82574185835055
``` |

**Code listing 2.1.** Computing the mean and standard deviation of two sets of numbers.



### 2.1.3  Variance

The variance is the square of the standard deviation; standard deviation is the positive square root of the variance. Thus they are computed in the same form, with `StandardDeviation()` replaced by `Variance()`.

### 2.1.4  Covariance

The covariance is a measure of relationship between two variables. The previous example datasets were all univariate: i.e. they only involved a single variable. When there are more than one variable, it is often useful to know how those variables are related. For example, suppose that when a given variable increases, another variable also increases. Then those variables are somehow related; we expect their covariance to be positive. When a variable increases and other decreases, their covariance would be negative.

Suppose we would like to compute the covariance between the two variables from the previous example. To do so we may use

```
double cov = x1.Covariance(x2);
```

revealing a value of 14.667.

### 2.1.5  Covariance Matrix

Remember that the covariance is a measure of relationship between **two** variables. However, in a typical data set we have much more than just two variables. Yet it is useful to know the relationship between all variables. In other words, we could compute the variance of all variable pairs and summarize this information in the form of a **covariance matrix**.

| X | Y | Z |
|---|---|---|
| … | … | … |
|   |   |   |
|   |   |   |
|   |   |   |

|   | X | Y | Z |
|---|---|---|---|
| X |   |   |   |
| Y |   |   |   |
| Z |   |   |   |

**Table 1.** Example data set and its corresponding covariance matrix on the right.



The covariance matrix is a square matrix with as many rows and columns as variables in your data. In the example above, there are three variables, X, Y, and Z. The corresponding covariance matrix, given on the right, has one row and one column for each of those three variables, and each element in the covariance matrix expresses how those variables change together. The diagonal elements of the covariance matrix contain the **variances** for each variable.

Now, suppose we were given the same data as Lindsay's (pg. 7). Let's express this data as a .NET matrix so we can process it using Accord.NET.

```
double[,] data =
{
    // Hours (H)   Mark (M)
    {     9,         39    },
    {    15,         56    },
    {    25,         93    },
    {    14,         61    },
    {    10,         50    },
    {    18,         75    },
    {     0,         32    },
    {    16,         85    },
    {     5,         42    },
    {    19,         70    },
    {    16,         66    },
    {    20,         80    }
};
```

After that, we can compute its covariance matrix by calling

```
double[,] covarianceMatrix = data.Covariance();
```

We can also quickly visualize the data by using

```
ScatterplotBox.Show(data);
```

The code listing on the next page summarizes how to replicate Lindsay's calculations using Accord.NET. The scatterplot generated by the ScatterplotBox command is shown in the sequence.



| Program | Output |
|---|---|
| ```
double[,] data =
{
    // Hours (H)   Mark (M)
    {     9,         39    },
    {    15,         56    },
    {    25,         93    },
    {    14,         61    },
    {    10,         50    },
    {    18,         75    },
    {     0,         32    },
    {    16,         85    },
    {     5,         42    },
    {    19,         70    },
    {    16,         66    },
    {    20,         80    }
};

// Compute total and average
double[] totals = data.Sum();
double[] averages = data.Mean();

// Compute covariance matrix
double[,] C = data.Covariance();

// Show results on screen
Console.WriteLine("Data: ");
Console.WriteLine("   Hours(H)  Mark(M)");
Console.WriteLine(data.ToString("     00"));

Console.WriteLine();
Console.WriteLine("Sum: " + totals.ToString("000.00"));
Console.WriteLine("Avg: " + averages.ToString(" 00.00"));

Console.WriteLine();
Console.WriteLine("Covariance matrix:");
Console.WriteLine(C.ToString(" 000.00"));

Console.ReadKey();
ScatterplotBox.Show(data);
``` | ```
Data:
   Hours(H)   Mark(M)
      09        39
      15        56
      25        93
      14        61
      10        50
      18        75
      00        32
      16        85
      05        42
      19        70
      16        66
      20        80

Sum: 167.00 749.00
Avg:  13.92  62.42

Covariance matrix:
 047.72   122.95
 122.95   370.08
```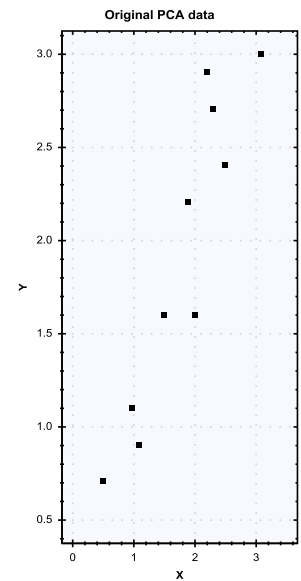 |

**Code listing 2.2.** Calculating a covariance matrix. The figure on the right shows the scatterplot generated using the ScatterplotBox class of the framework.

## 2.2 Matrix algebra

Matrix algebra in the Accord.NET Framework can be performed using extension methods as well. Table 2 summarizes common matrix operations and their counterparts using Octave [**14**].

Table 2. Basic matrix operations

| Operation | Accord.NET (C#) | Octave |
|---|---|---|
| Transpose | var At = A.Transpose(); | At = A' |
| Inverse | var invA = A.Inverse(); | invA = inv(A) |
| Pseudo-Inverse | var pinvA = A.PseudoInverse(); | pinvA = pinv(A) |

Table 3. Basic matrix algebra

| Operation | Accord.NET (C#) | Octave |
|---|---|---|
| Addition | var C = A.Add(B); | C = A + B |
| Subtraction | var C = A.Subtract(B); | C = A - B |
| Multiplication | var C = A.Multiply(B); | C = A * B |
| Division | var C = A.Divide(B); | C = A / B |

Table 4. Basic elementwise operations

| Operation | Accord.NET (C#) | Octave |
|---|---|---|
| Multiplication | var C = A.ElementwiseMultiply(B); | C = A .* B |
| Division | var C = A.ElementwiseDivide(B); | C = A ./ B |
| Power | var C = A.ElementwisePower(B); | C = A .^ B |

Consider now the example for matrix-vector multiplication given by Lindsay on page 9. This example shows how to multiply vectors and matrices, as well as demonstrates what characterizes an eigenvector. Eigenvectors and eigenvalues are discussed in more details in the next sections.

| Program | Output |
|---|---|
| ```csharp
// Consider the following matrix
double[,] A =
{
    { 2, 3 },
    { 2, 1 }
};

// Now consider the vector
double[] u = { 1, 3 };

// Multiplying both, we get x = [ 11 5 ]'
double[] x = A.Multiply(u);

// We can not express 'x' as a multiple
// of 'u', so 'u' is not an eigenvector.

// However, consider now the vector
double[] v = { 3, 2 };

// Multiplying both, we get y = [ 12 8 ]'
double[] y = A.Multiply(v);

// It can be seen that 'y' can be expressed as
// a multiple of 'v'. Since y = 4*v, 'v' is an
// eigenvector with the associated eigenvalue 4.

// Show on screen
Console.WriteLine("Matrix A:");
Console.WriteLine(A.ToString(" 0"));

Console.WriteLine();
Console.WriteLine("Vector u:");
Console.WriteLine(u.Transpose().ToString(" 0"));

Console.WriteLine();
Console.WriteLine("Vector v:");
Console.WriteLine(v.Transpose().ToString(" 0"));

Console.WriteLine();
Console.WriteLine("x = A*u");
Console.WriteLine(x.Transpose().ToString(" 0"));

Console.WriteLine();
Console.WriteLine("y = A*v");
Console.WriteLine(y.Transpose().ToString(" 0"));
``` | Matrix A:<br> 2  3<br> 2  1<br><br>Vector u:<br> 1<br> 3<br><br>Vector v:<br> 3<br> 2<br><br>x = A*u<br> 11<br>  5<br><br>y = A*v<br> 12<br>  8 |

**Code listing 2.3.** Matrix vector multiplication.



## 2.2.1 Eigenvectors, Eigenvalues and the Eigendecomposition

For a $n \times n$ square matrix $\boldsymbol{A}$, a vector $\boldsymbol{v}$ of length $n$ is said to be an eigenvector of $\boldsymbol{A}$ if and only if it satisfies the equation

$$\boldsymbol{Av} = \lambda \boldsymbol{v}.$$

In the previous equation, $\lambda$ is an scalar; a real number such that, when multiplied with $v$, gives the same result as if the original matrix was multiplied with $v$. When we succeed in finding one eigenvector $v$ for the matrix $A$, its associated scalar value $\lambda$ will be called the *eigenvalue associated with the eigenvector $v$*.

The process of finding the eigenvalues and eigenvectors of a matrix is known as the *Eigenvalue decomposition*. The Eigenvalue decomposition (also known as eigendecomposition or EVD for short) is the process of factorizing a matrix such that this matrix can be written in terms of two other matrices $\boldsymbol{V}$ and $\boldsymbol{\Lambda}$, in which $\boldsymbol{V}$ will be a matrix containing all eigenvectors along its columns and $\boldsymbol{\Lambda}$ is a diagonal matrix whose diagonal elements will be the eigenvalues of $\boldsymbol{A}$:

$$\boldsymbol{A} = \boldsymbol{V} \boldsymbol{\Lambda} \boldsymbol{V}^{-1}$$

If $\boldsymbol{A}$ is symmetric and we choose $\boldsymbol{V}$ to be orthonormal, then the expression simplifies to

$$\boldsymbol{A} = \boldsymbol{V} \boldsymbol{\Lambda} \boldsymbol{V}^{t}$$

This will be useful in the next chapter as we will see that, for performing PCA one computes the Eigendecomposition of a Covariance matrix. As shown in section 2.1.5, covariance matrices are symmetric by definition. Because of this, we will not be giving more details for when A is arbitrary. An extraordinarily and complete book is due to Kenneth Kuttler and is freely available from [15].

For an example, consider the symmetric matrix

$$\boldsymbol{M} = \begin{pmatrix} 3 & 2 & 4 \\ 2 & 0 & 2 \\ 4 & 2 & 3 \end{pmatrix}$$



Now, check what happens when we multiply $M$ by the vector $v = (-1, 0, 1)^t$

$$M \begin{pmatrix} -1 \\ 0 \\ 1 \end{pmatrix} = \begin{pmatrix} 3 & 2 & 4 \\ 2 & 0 & 2 \\ 4 & 2 & 3 \end{pmatrix} \begin{pmatrix} -1 \\ 0 \\ 1 \end{pmatrix} = \begin{pmatrix} 1 \\ 0 \\ -1 \end{pmatrix}$$

The resulting vector $(1, 0, -1)^t$ is clearly a scaled version of $v$ – it is, in fact, its exact opposite – obtained by taking $v$ and multiplying with the scalar $-1$. They form, thus, our first eigenvector-eigenvalue pair $(\lambda_1, v_1)$ given by taking $\lambda_1 = -1$ and $v_1 = (-1, 0, 1)^t$.

Of course, eigenvectors are not detected by guessing. There are analytical forms to obtain the matrix of eigenvectors $V$ and the diagonal matrix of eigenvalues $\Lambda$. While those methods are out of scope for this tutorial, the code listing in the next page demonstrates how to compute the Eigendecomposition using the framework.

The Eigenvalue decomposition has many interesting applications. Since the original matrix can be decomposed into a diagonal matrix (so in fact a set of scalar elements) and an **orthogonal** matrix, one can use the neat properties of orthogonal matrices to form a basis for other coordinate systems. In fact, this is exactly the application case for PCA: by forming an alternate coordinate system in which each variable is decorrelated from each other, it is possible to detect dependencies among variables and *discard potentially[1] unneeded data.*

---

[1] It all depends on the application – never assume projections given by a decorrelated basis to be optimal for all applications. Often the discarded information may still be important, if not critical, such as in the case for classification. See [**17**] for an example.



| Program | Output |
|---|---|
| ```csharp
// Consider the following matrix
double[,] M =
{
    {  3,  2,  4 },
    {  2,  0,  2 },
    {  4,  2,  3 }
};

// Create an Eigenvalue decomposition
var evd = new EigenvalueDecomposition(M);

// Store the eigenvalues and eigenvectors
double[] λ = evd.RealEigenvalues;
double[,] V = evd.Eigenvectors;

// Reconstruct M = V*λ*V'
double[,] R =
V.MultiplyByDiagonal(λ).MultiplyByTranspose(V);

// Show on screen
Console.WriteLine("Matrix: ");
Console.WriteLine(M.ToString(" 0"));

Console.WriteLine();
Console.WriteLine("Eigenvalues: ");
Console.WriteLine(λ.ToString(" +0.000; -0.000;"));

Console.WriteLine();
Console.WriteLine("Eigenvectors:");
Console.WriteLine(V.ToString(" +0.000; -0.000;"));

Console.WriteLine();
Console.WriteLine("Reconstruction:");
Console.WriteLine(R.ToString(" 0"));
``` | ```
Matrix:
 3  2  4
 2  0  2
 4  2  3

Eigenvalues:
 -1.000  -1.000  +8.000

Eigenvectors:
 -0.041  -0.744  +0.667
 +0.910  +0.248  +0.333
 -0.413  +0.620  +0.667

Reconstruction:
 3  2  4
 2  0  2
 4  2  3
``` |

**Code listing 2.4.** Computing the Eigenvalue decomposition with Accord.NET.

# 3 Principal Component Analysis

Principal Component Analysis, or PCA, is just another name for projecting the data into the first orthogonal vectors obtained by Eigendecomposing its covariance matrix.

Recall that the covariance matrix summarizes the relationships between all variables in the data. Strongly related variables may result in a covariance matrix with large values outside its main diagonal. However, one may see highly dependent variables as redundant: This is the case when we have the same data represented in different units – such as same data recorded twice, but in one column registered in Kilograms and in other column in Pounds. For most practical purposes, this would be redundant information and could be safely discarded without any loss of information.

In this case, what one would like to do is to remove those large values from outside the main diagonal of the covariance matrix. To do so, one can compute the Eigendecomposition of the covariance matrix to form a new, uncorrelated basis for the data. After projecting the original data into this new basis, the data becomes uncorrelated.

The magnitude of the eigenvalues gives a measure of importance for each of the eigenvectors forming the base. In the case of the exact same data represented using different units of measurement, one of the eigenvalues will surely be zero. The eigenvector associated with this zero eigenvalue could safely be dropped from the basis, causing dimensionality reduction.

There are, however, many ways to achieve this decorrelation. First we will reproduce the same steps given by Lindsay on her tutorial, and then show how we can achieve more stable results using the Singular value decomposition of the original data matrix instead of the Eigendecomposition of the covariance matrix. All steps are computed using the exact same data given by Lindsay's page 12.

## 3.1 The Eigendecomposition Method

**Step 1.** Get some data

```
double[,] data =
{
    { 2.5,  2.4 },
    { 0.5,  0.7 },
    { 2.2,  2.9 },
    { 1.9,  2.2 },
    { 3.1,  3.0 },
    { 2.3,  2.7 },
    { 2.0,  1.6 },
    { 1.0,  1.1 },
    { 1.5,  1.6 },
    { 1.1,  0.9 }
};
```

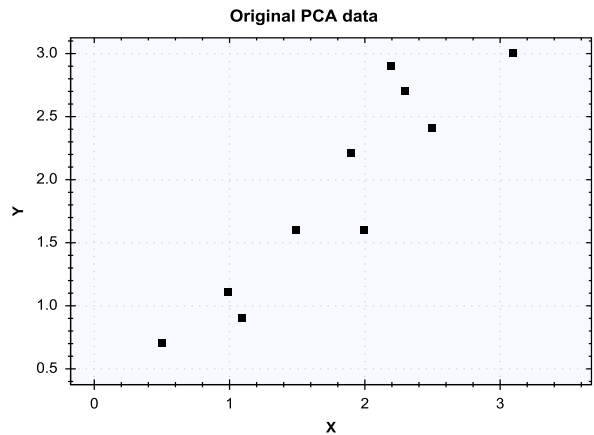

**Figure 2.** Plot of the data

**Step 2.** Subtract the mean.

```
double[] mean = data.Mean();
double[,] dataAdjust = data.Subtract(mean);
```

After this step, dataAdjust should contain the same values as given in Lindsay tutorial, pg. 13.

|        | x   | y   |              | x     | y     |
|--------|-----|-----|--------------|-------|-------|
|        | 2.5 | 2.4 |              | 0.69  | 0.49  |
|        | 0.5 | 0.7 |              | -1.31 | -1.21 |
|        | 2.2 | 2.9 |              | 0.39  | 0.99  |
|        | 1.9 | 2.2 |              | 0.09  | 0.29  |
| Data = | 3.1 | 3.0 | dataAdjust = | 1.29  | 1.09  |
|        | 2.3 | 2.7 |              | 0.49  | 0.79  |
|        | 2.0 | 1.6 |              | 0.19  | -0.31 |
|        | 1.0 | 1.1 |              | -0.81 | -0.81 |
|        | 1.5 | 1.6 |              | -0.31 | -0.31 |
|        | 1.1 | 0.9 |              | -0.71 | -1.01 |

**Step 3.** Calculate the covariance matrix.

```
double[,] cov = dataAdjust.Covariance();
```

After this step, the matrix `cov` should contain the values

$$cov = \begin{pmatrix} 0.6165555556 & 0.6154444444 \\ 0.6154444444 & 0.7165555556 \end{pmatrix}$$

**Step 4.** Calculate the Eigenvectors of the covariance matrix

```
var evd = new EigenvalueDecomposition(cov);

double[] eigenvalues = evd.RealEigenvalues;
double[,] eigenvectors = evd.Eigenvectors;

// Sort eigenvalues and vectors in descending order
eigenvectors = Matrix.Sort(eigenvalues, eigenvectors,
  new GeneralComparer(ComparerDirection.Descending, true));
```

At this point, the eigenvalues vector and the eigenvectors matrix should contain the values

$$eigenvalues = \begin{pmatrix} 1.2840277122 \\ 0.0490833989 \end{pmatrix}$$

$$eigenvectors = \begin{pmatrix} 0.6778733985 & -0.7351786555 \\ 0.7351786555 & 0.6778733985 \end{pmatrix}$$

The corresponding eigenvalue-eigenvector pairs can be identified by their matching colors. The aware reader may note that the eigenvalues are swapped when compared to the eigenvalues given by Lindsay. This is because we have used the Matrix.Sort method to sort the eigenvalues (and corresponding eigenvectors) according to their magnitude, so larger eigenvalues come first.

The even more aware reader may note that the second eigenvector (in red) has opposite signs when compared to Lindsay's first eigenvector. This is discussed in Lindsay's tutorial: Any multiple of an eigenvalue is still an eigenvalue, so Accord.NET may produce results with different signs than other software packages. This is expected, as the Eigendecomposition is not unique.

**Step 5.** Choosing components and forming a feature vector

Here we can either choose all eigenvectors to form a new basis...

```
double[,] featureVector = eigenvectors;
```

... or, instead, select only the first eigenvector.

```
double[,] featureVector = eigenvectors.GetColumn(0).Transpose();
```

**Step 6.** Derive the new data set

```
double[,] finalData = dataAdjust.Multiply(eigenvectors);
```

After this step, finalData should contain the same values reported by Lindsay, except for the inverted sign of the first column. Again, this is expected; what matters is that both columns have been decorrelated.

$$\text{finalData} = \begin{array}{|cc|} \hline 1^{st}\ \text{PC} & 2^{nd}\ \text{PC} \\ \hline 0.8279701862 & -0.1751153070 \\ -1.7775803253 & 0.1428572265 \\ 0.9921974944 & 0.3843749889 \\ 0.2742104160 & 0.1304172066 \\ 1.6758014186 & -0.2094984613 \\ 0.9129491032 & 0.1752824436 \\ -0.0991094375 & -0.3498246981 \\ -1.1445721638 & 0.0464172582 \\ -0.4380461368 & 0.0177646297 \\ -1.2238205551 & -0.1626752871 \\ \hline \end{array}$$

If, instead, we had chosen to discard the last vector, our finalData would be

$$\text{finalData} = \begin{array}{|c|} \hline 1^{st}\ \text{PC} \\ \hline 0.8279701862 \\ -1.7775803253 \\ 0.9921974944 \\ 0.2742104160 \\ 1.6758014186 \\ 0.9129491032 \\ -0.0991094375 \\ -1.1445721638 \\ -0.4380461368 \\ -1.2238205551 \\ \hline \end{array}$$

| Program | Output |
|---|---|

```csharp
// Step 1. Get some data
double[,] data =
{
    { 2.5,  2.4 },
    { 0.5,  0.7 },
    { 2.2,  2.9 },
    { 1.9,  2.2 },
    { 3.1,  3.0 },
    { 2.3,  2.7 },
    { 2.0,  1.6 },
    { 1.0,  1.1 },
    { 1.5,  1.6 },
    { 1.1,  0.9 }
};

// Step 2. Subtract the mean
double[] mean = data.Mean();
double[,] dataAdjust = data.Subtract(mean);

// Step 3. Calculate the covariance matrix
double[,] cov = dataAdjust.Covariance();

// Step 4. Calculate the eigenvectors and
//  eigenvalues of the covariance matrix
var evd = new EigenvalueDecomposition(cov);

double[] eigenvalues = evd.RealEigenvalues;
double[,] eigenvectors = evd.Eigenvectors;

// Step 5. Choosing components and
//         forming a feature vector

// Sort eigenvalues and vectors in descending order
eigenvectors = Matrix.Sort(eigenvalues, eigenvectors,
 new GeneralComparer(ComparerDirection.Descending, true));

// Select all eigenvectors
double[,] featureVector = eigenvectors;

// Step 6. Deriving the new data set
double[,] finalData = dataAdjust.Multiply(eigenvectors);
```

```
Data

   x       y
 ------------
  2.5     2.4
  0.5     0.7
  2.2     2.9
  1.9     2.2
  3.1     3.0
  2.3     2.7
  2.0     1.6
  1.0     1.1
  1.5     1.6
  1.1     0.9

Data Adjust

   x       y
 ------------
  0.69    0.49
 -1.31   -1.21
  0.39    0.99
  0.09    0.29
  1.29    1.09
  0.49    0.79
  0.19   -0.31
 -0.81   -0.81
 -0.31   -0.31
 -0.71   -1.01

Covariance Matrix:
  +0.6165555556   +0.6154444444
  +0.6154444444   +0.7165555556

Eigenvalues:
  +1.2840277122   +0.0490833989

Eigenvectors:
  +0.6778733985   -0.7351786555
  +0.7351786555   +0.6778733985

Transformed Data

    x              y
 ---------------------------
  0.8279701862  -0.1751153070
 -1.7775803253   0.1428572265
  0.9921974944   0.3843749889
  0.2742104160   0.1304172066
  1.6758014186  -0.2094984613
  0.9129491032   0.1752824436
 -0.0991094375  -0.3498246981
 -1.1445721638   0.0464172582
 -0.4380461368   0.0177646297
 -1.2238205551  -0.1626752871
```

```
// Show on screen
Console.WriteLine("Data");
Console.WriteLine();
Console.WriteLine("    x       y");
Console.WriteLine(" ------------");
Console.WriteLine(data.ToString("  0.0 "));

Console.ReadKey();

Console.WriteLine();
Console.WriteLine("Data Adjust");
Console.WriteLine();
Console.WriteLine("    x       y");
Console.WriteLine(" ------------");
Console.WriteLine(dataAdjust.ToString("  0.00; -0.00;"));

Console.ReadKey();
ScatterplotBox.Show("Original PCA data", data);

Console.ReadKey();
Console.WriteLine();
Console.WriteLine("Covariance Matrix: ");
Console.WriteLine(cov.ToString(" +0.0000000000; -0.0000000000;"));

Console.WriteLine();
Console.WriteLine("Eigenvalues: ");
Console.WriteLine(eigenvalues.ToString(" +0.0000000000; -0.0000000000;"));

Console.WriteLine();
Console.WriteLine("Eigenvectors:");
Console.WriteLine(eigenvectors.ToString(" +0.0000000000; -0.0000000000;"));

Console.ReadKey();

Console.WriteLine();
Console.WriteLine("Transformed Data");
Console.WriteLine();
Console.WriteLine("    x                y");
Console.WriteLine(" ---------------------------");

Console.WriteLine(finalData.ToString("  0.0000000000; -0.0000000000;"));

ScatterplotBox.Show("Transformed PCA data", finalData);
```

Code listing 3.1. Principal Component Analysis using the Covariance method.

## 3.2 The Singular Value Decomposition Method

Now, here is the alternate version used internally by Accord.NET. Instead of calculating the covariance matrix for the data and then Eigendecomposing it, a more stable alternative is to compute the Singular value decomposition of the data matrix directly.

Given a rectangular $m \times n$ matrix $\boldsymbol{A}$, the Singular Value Decomposition (SVD) decomposes $\boldsymbol{A}$ into three matrices $\boldsymbol{U}$, $\boldsymbol{\Sigma}$, $\boldsymbol{V}$ such that

$$\boldsymbol{A} = \boldsymbol{U}\,\boldsymbol{\Sigma}\,\boldsymbol{V}^t$$

In which $\boldsymbol{U}$ and $\boldsymbol{V}$ are orthogonal matrices containing the left and right singular vectors, respectively, and $\boldsymbol{\Sigma}$ is a diagonal matrix containing the singular values on its diagonal.

Recall now that the Eigendecomposition method for PCA computes the Eigendecomposition of the covariance matrix. If $\boldsymbol{A}$ holds our centered data matrix, the sample covariance matrix can be written in matrix notation as

$$\boldsymbol{C} = \frac{1}{n-1}\,\boldsymbol{A}^t\boldsymbol{A}.$$

It is possible, thus, to express $\boldsymbol{C}$ in terms of the scatter matrix[1] $\boldsymbol{A}^t\boldsymbol{A}$. A drawback of computing the covariance matrix is that it may result in loss of precision and numerical instabilities (e.g. the *Läuchli* matrix). A better approach would be to compute the orthogonal basis from the data matrix $\boldsymbol{A}$ directly. Consider now what happens when one attempts to identify the Eigenvalue decomposition of $\boldsymbol{A}^t\boldsymbol{A}$:

$$\boldsymbol{A}^t\boldsymbol{A} = \boldsymbol{V}\,\boldsymbol{\Lambda}\,\boldsymbol{V}^t$$

Using the *Cholesky* decomposition of $\boldsymbol{\Lambda} = \boldsymbol{\Sigma}^t\boldsymbol{\Sigma}$, one can write

$$\boldsymbol{A}^t\boldsymbol{A} = \boldsymbol{V}\,\boldsymbol{\Lambda}\,\boldsymbol{V}^t = \boldsymbol{V}\,(\boldsymbol{\Sigma}^t\boldsymbol{\Sigma})\,\boldsymbol{V}^t$$

---

[1] In statistics, the matrix $\boldsymbol{A}^t\boldsymbol{A}$ is also known as the scatter matrix. It can be seen as an unnormalized version of the covariance matrix.

Then, considering an orthonormal matrix $\mathbf{U}$, we can use the identity $\mathbf{I} = \mathbf{U^t U}$ such that

$$\mathbf{A^t A} = \mathbf{V \Lambda V^t} = \mathbf{V \Sigma^t \Sigma V^t} = \mathbf{V \Sigma^t I \Sigma V^t} = \mathbf{V \Sigma^t U^t U \Sigma V^t}$$

And finally, by noticing that the SVD of $\mathbf{A^t}$ and $\mathbf{A}$ can be written as, respectively, $\mathbf{A^t} = \mathbf{V \Sigma^t U^t}$ and $\mathbf{A} = \mathbf{U \Sigma V^t}$, one finally arrives at the relations

$$(\mathbf{V \Sigma^t U^t})(\mathbf{U \Sigma V^t}) = \mathbf{A^t A}$$

$$(\mathbf{U \Sigma V^t})(\mathbf{V \Sigma^t U^t}) = \mathbf{A A^t}$$

Allowing one to conclude that

- The right singular vectors $\mathbf{V}$ can be seen as the eigenvectors of $\mathbf{A^t A}$
- The left singular vectors $\mathbf{U}$ can be seen as the eigenvectors of $\mathbf{A A^t}$
- And that the non-zero diagonal elements values of $\mathbf{\Sigma}$ will be the square roots of the eigenvalues of either $\mathbf{A^t A}$ or $\mathbf{A A^t}$.

Using the aforementioned relations, one can express PCA in terms of the Singular Value Decomposition in the following way:

**Step 1.** Get some data

```
double[,] data =
{
    { 2.5,  2.4 },
    { 0.5,  0.7 },
    { 2.2,  2.9 },
    { 1.9,  2.2 },
    { 3.1,  3.0 },
    { 2.3,  2.7 },
    { 2.0,  1.6 },
    { 1.0,  1.1 },
    { 1.5,  1.6 },
    { 1.1,  0.9 }
};
```

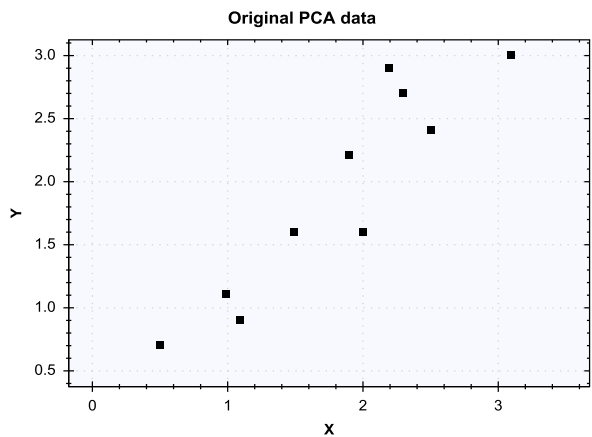

**Step 2.** Subtract the mean.

```
double[] mean = data.Mean();
double[,] dataAdjust = data.Subtract(mean);
```

After this step, dataAdjust should contain the same values as given in Lindsay tutorial, pg. 13.

|  | x | y |
|---|---|---|
|  | 2.5 | 2.4 |
|  | 0.5 | 0.7 |
|  | 2.2 | 2.9 |
|  | 1.9 | 2.2 |
| Data = | 3.1 | 3.0 |
|  | 2.3 | 2.7 |
|  | 2.0 | 1.6 |
|  | 1.0 | 1.1 |
|  | 1.5 | 1.6 |
|  | 1.1 | 0.9 |

|  | x | y |
|---|---|---|
|  | 0.69 | 0.49 |
|  | -1.31 | -1.21 |
|  | 0.39 | 0.99 |
|  | 0.09 | 0.29 |
| dataAdjust = | 1.29 | 1.09 |
|  | 0.49 | 0.79 |
|  | 0.19 | -0.31 |
|  | -0.81 | -0.81 |
|  | -0.31 | -0.31 |
|  | -0.71 | -1.01 |

**Step 3.** Calculate the SVD of the adjusted data matrix

```
var svd = new SingularValueDecomposition(dataAdjust);

double[] singularValues = svd.Diagonal;
double[,] eigenvectors = svd.RightSingularVectors;
```

At this point, the *singularValues* vector and the *eigenvectors* matrix should contain the values

$$singularValues = \begin{pmatrix} 3.3994483978 \\ 0.6646432054 \end{pmatrix}$$

$$eigenvectors = \begin{pmatrix} 0.6778733985 & -0.7351786555 \\ 0.7351786555 & 0.6778733985 \end{pmatrix}$$

**Step 4.** Calculate the eigenvalues as the squares of the singular values

```
double[] eigenvalues = singularValues.ElementwisePower(2);
```

However, remember that SVD computes the *Eigendecomposition* of the scatter matrix $A^t A$, and not of the covariance matrix $A^t A \frac{1}{n-1}$. So we still have to divide the eigenvalues by $n-1$:

```
eigenvalues = eigenvalues.Divide(data.GetLength(0) - 1);
```

Resulting in

$$eigenvalues = \begin{pmatrix} 1.2840277122 \\ 0.0490833989 \end{pmatrix}$$

which are the same values given by Lindsay. After that, all subsequent steps are the same and need to be shown. The code listing on the next page summarizes all steps required to perform PCA using the SVD method.

| Program | Output |
|---|---|
| | |

```
// Step 1. Get some data
double[,] data =
{
    { 2.5,  2.4 },
    { 0.5,  0.7 },
    { 2.2,  2.9 },
    { 1.9,  2.2 },
    { 3.1,  3.0 },
    { 2.3,  2.7 },
    { 2.0,  1.6 },
    { 1.0,  1.1 },
    { 1.5,  1.6 },
    { 1.1,  0.9 }
};

// Step 2. Subtract the mean
double[] mean = data.Mean();
double[,] dataAdjust = data.Subtract(mean);

// Step 3. Calculate the singular values and
//   singular vectors of the data matrix
var svd = new SingularValueDecomposition(dataAdjust);

double[] singularValues = svd.Diagonal;
double[,] eigenvectors = svd.RightSingularVectors;

// Step 4. Calculate the eigenvalues as
//   the square of the singular values
double[] eigenvalues = singularValues.ElementwisePower(2);

// Step 5. Choosing components and
//         forming a feature vector

// Select all eigenvectors
double[,] featureVector = eigenvectors;

// Step 6. Deriving the new data set
double[,] finalData = dataAdjust.Multiply(eigenvectors);
```

```
Data

    x      y
  -----------
   2.5    2.4
   0.5    0.7
   2.2    2.9
   1.9    2.2
   3.1    3.0
   2.3    2.7
   2.0    1.6
   1.0    1.1
   1.5    1.6
   1.1    0.9

Data Adjust

    x      y
  -----------
   0.69   0.49
  -1.31  -1.21
   0.39   0.99
   0.09   0.29
   1.29   1.09
   0.49   0.79
   0.19  -0.31
  -0.81  -0.81
  -0.31  -0.31
  -0.71  -1.01

Singular values:
 +3.3994483978  +0.6646432054

Eigenvalues:
 +11.5562494096  +0.4417505904

Eigenvalues (normalized):
 +1.2840277122  +0.0490833989

Eigenvectors:
 +0.6778733985  -0.7351786555
 +0.7351786555  +0.6778733985

Transformed Data

     x              y
  ---------------------------
  +0.8279701862  -0.1751153070
  -1.7775803253  +0.1428572265
  +0.9921974944  +0.3843749889
  +0.2742104160  +0.1304172066
  +1.6758014186  -0.2094984613
  +0.9129491032  +0.1752824436
  -0.0991094375  -0.3498246981
  -1.1445721638  +0.0464172582
  -0.4380461368  +0.0177646297
  -1.2238205551  -0.1626752871
```

```csharp
// Show on screen.
Console.WriteLine("Data");
Console.WriteLine();
Console.WriteLine("   x      y");
Console.WriteLine(" ------------");
Console.WriteLine(data.ToString("  0.0 "));

Console.ReadKey();

Console.WriteLine();
Console.WriteLine("Data Adjust");
Console.WriteLine();
Console.WriteLine("   x      y");
Console.WriteLine(" ------------");
Console.WriteLine(dataAdjust.ToString("  0.00; -0.00;"));

Console.ReadKey();
ScatterplotBox.Show("Original PCA data", data);

Console.WriteLine();
Console.WriteLine("Singular values: ");
Console.WriteLine(singularValues.ToString(" +0.0000000000; -0.0000000000;"));

Console.WriteLine();
Console.WriteLine("Eigenvalues: ");
Console.WriteLine(eigenvalues.ToString(" +0.0000000000; -0.0000000000;"));

// Normalize eigenvalues to replicate the covariance
eigenvalues = eigenvalues.Divide(data.GetLength(0) - 1);

Console.WriteLine();
Console.WriteLine("Eigenvalues (normalized): ");
Console.WriteLine(eigenvalues.ToString(" +0.0000000000; -0.0000000000;"));

Console.WriteLine();
Console.WriteLine("Eigenvectors:");
Console.WriteLine(eigenvectors.ToString(" +0.0000000000; -0.0000000000;"));

Console.ReadKey();

Console.WriteLine();
Console.WriteLine("Transformed Data");
Console.WriteLine();
Console.WriteLine("   x                y");
Console.WriteLine(" ---------------------------");
Console.WriteLine(finalData.ToString(" +0.0000000000; -0.0000000000;"));
```

Code listing 3.2. Principal Component Analysis using the SVD method.

## 3.3 The Accord.NET Framework Method

This section shows how to use the built-in *PrincipalComponentAnalysis* class from the Accord.NET Framework to compute PCA automatically. There are very few steps in computing the analysis this way; all it involves is creating a *PrincipalComponentAnalysis* object and calling its *Compute* method.

To create a PrincipalComponentAnalysis object, one can use

```
var pca = new PrincipalComponentAnalysis(data);
```

It is then possible to set a few settings about the analysis. For example, the analysis could be set to conserve memory by performing operations in-place, rewriting the original data matrix:

```
pca.Overwrite = true;
```

Or it could also be set to use the analysis by *correlation*, which is more indicated when analysing data with highly different measurement units:

```
pca.Method = AnalysisMethod.Standardize;
```

The next step is to finally compute the analysis. This involves a single method call to

```
pca.Compute();
```

After the analysis is computed, many results and measures become available through the analysis object. Many of them can be directly data-bound into visual controls as demonstrated in the Accord.NET's PCA sample application. For example, one can get access to the individual importance of each component by accessing the *ComponentProportions* property.

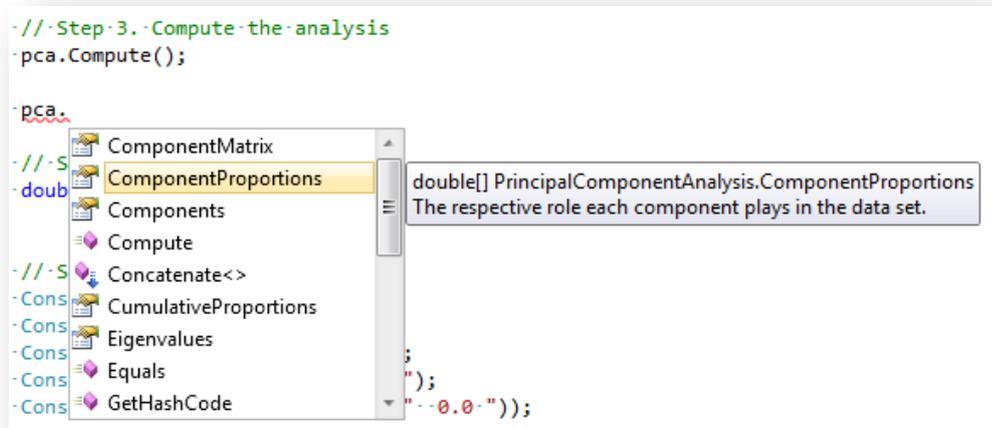

Or one could also get detailed information about each principal component, such as their related eigenvalue, related eigenvector, proportions and more by navigating in the *Components* property.

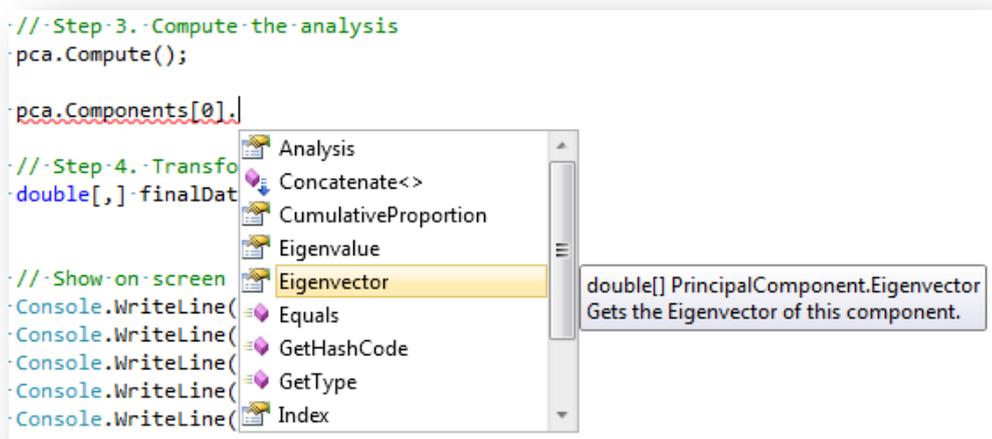

One interesting characteristic of the PrincipalComponent objects is that they can be directly databound to visual controls, such as a Windows Forms DataGridView:

```
dataGridView1.DataSource = pca.Components;
```

which leads to results similar to the ones shown by the Accord.NET Framework's PCA Sample Application. All sample applications (including source code) comes together with the framework executable installer and compressed archive. They are also available as standalone downloads from the project's site. For more details, please visit the project's web page on Google Code at http://accord.googlecode.com.

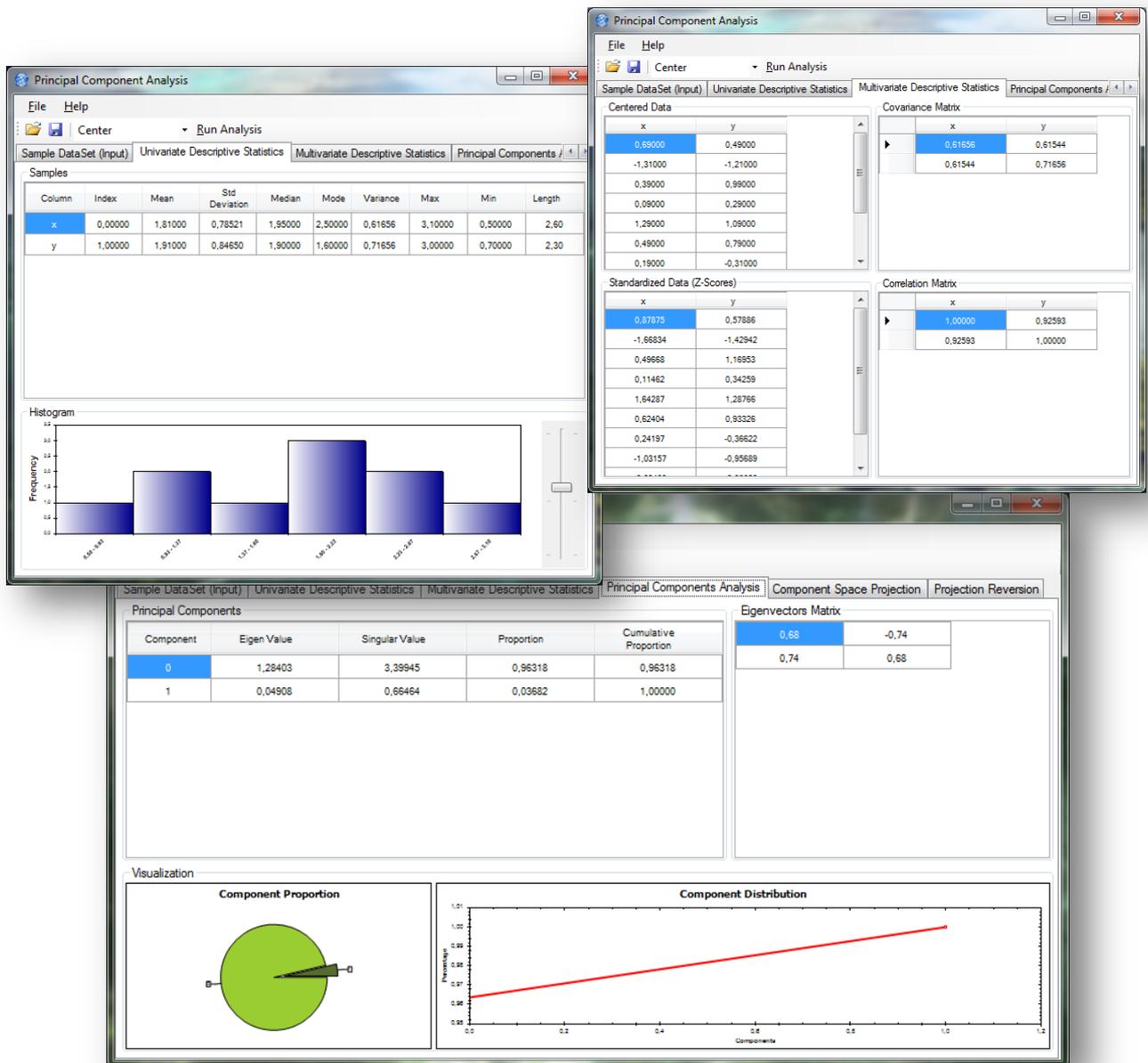

**Figure 3.** Accord.NET's Principal Component Analysis sample application

| Program | Output |
|---|---|
| ```csharp
// Step 1. Get some data
double[,] data =
{
    { 2.5,  2.4 },
    { 0.5,  0.7 },
    { 2.2,  2.9 },
    { 1.9,  2.2 },
    { 3.1,  3.0 },
    { 2.3,  2.7 },
    { 2.0,  1.6 },
    { 1.0,  1.1 },
    { 1.5,  1.6 },
    { 1.1,  0.9 }
};

// Step 2. Create the Principal Component Analysis
var pca = new PrincipalComponentAnalysis(data);

// Step 3. Compute the analysis
pca.Compute();

// Step 4. Transform your data
double[,] finalData = pca.Transform(data);

// Show on screen
Console.WriteLine("Data");
Console.WriteLine();
Console.WriteLine("    x       y");
Console.WriteLine(" ------------");
Console.WriteLine(data.ToString("  0.0 "));
Console.ReadKey();
ScatterplotBox.Show("Original PCA data", data);
Console.WriteLine();
Console.WriteLine("Eigenvalues: ");
Console.WriteLine(pca.Eigenvalues
  .ToString(" +0.0000000000; -0.0000000000;"));
Console.WriteLine();
Console.WriteLine("Eigenvectors:");
Console.WriteLine(pca.ComponentMatrix
  .ToString(" +0.0000000000; -0.0000000000;"));
Console.ReadKey();
Console.WriteLine();
Console.WriteLine("Transformed Data");
Console.WriteLine();
Console.WriteLine("       x             y");
Console.WriteLine(" ---------------------------");
Console.WriteLine(finalData
  .ToString("  0.0000000000; -0.0000000000;"));
``` | ```
Data

   x        y
 ------------
  2.5      2.4
  0.5      0.7
  2.2      2.9
  1.9      2.2
  3.1      3.0
  2.3      2.7
  2.0      1.6
  1.0      1.1
  1.5      1.6
  1.1      0.9

Eigenvalues:
 +1.2840277122   +0.0490833989

Eigenvectors:
 +0.6778733985   -0.7351786555
 +0.7351786555   +0.6778733985

Transformed Data

       x              y
 ---------------------------
   0.8279701862   -0.1751153070
  -1.7775803253    0.1428572265
   0.9921974944    0.3843749889
   0.2742104160    0.1304172066
   1.6758014186   -0.2094984613
   0.9129491032    0.1752824436
  -0.0991094375   -0.3498246981
  -1.1445721638    0.0464172582
  -0.4380461368    0.0177646297
  -1.2238205551   -0.1626752871
``` |

**Code listing 3.3.** Calculating Principal Component Analysis with Accord.NET.



# 4 Conclusion

This technical report aimed to describe Principal Component Analysis and to provide answers to some common questions asked when performing such analysis with the Accord.NET Framework. While most of this tutorial is heavily based on Lindsay's excellent tutorial [1], we presented and discussed the alternate method using the Singular Value Decomposition instead of the Eigendecomposition of the covariance matrix. Furthermore, we have also left some interesting details such as differences between covariance and correlation anaylsis. Those may be addressed in future versions of this work.

As stated in the first page, this is a live document, and any suggestions or questions can either be sent by e-mail or asked in the Accord.NET Forums at http://groups.google.com/group/accord-net .

**Acknowledgements**. The author acknowledges Andrew Kirillov from the AForge.NET Framework [16] for providing such a wonderful base project to work upon and also the high number of both named and anonymous contributors who shared their suggestions, opinions and highly appreciated bug reports along those years.